\documentclass[twocolumn,shortnote]{jpsj2} 
\title{Coarse-Grained Picture for Controlling Complex Quantum Systems}

\author{Toshiya \textsc{Takami}$^{1}$\thanks{Present address: Computing and Communications Center, Kyushu University, Fukuoka, 812-8581, Japan.}
and Hiroshi \textsc{Fujisaki}$^{2}$}

\inst{$^{1}$
Institute for Molecular Science, Myodaiji, Okazaki, 444-8585, Japan \\
$^{2}$
Department of Chemistry,
Boston University, 590 Commonwealth Ave.,
Boston, Massachusetts, 02215, USA}

\abst{}

\kword{random matrix, quantum chaos, coarse-grain,
optimal control, rotating-wave approximation}

\begin{document}
\maketitle

Controlling atomic and molecular processes by laser fields is
one of current topics in physics and chemistry, and
there are various control schemes applied to such processes \cite{RZ00}.
These strategies are known to work when the 
system to be controlled is rather simple or small.
However, in reality, the system can become ``complex,''
where the dynamics will be described by 
multi-level-multi-level transitions with a random interaction. 
Although the laser field can be obtained by optimal control 
theory (OCT) \cite{RZ00} even for such complex systems,
it is difficult to analyze the controlled dynamics
because such a field obtained numerically
is often too complicated to interpret.
Hence it is desirable to have a more analytical point of view.

It is well known that the $\pi$ pulse\cite{AE87} or its generalizations 
can be employed to control few-level problems \cite{TN98}.
Recently, an analytic result for multi-level control problems
between general quantum states has been reported \cite{KAS02}.
The scheme is based on STIRAP \cite{RZ00}, 
and assumes an intermediate state coupled to the initial 
and target states.
Though the scheme can accomplish perfect control,
it relies on the energy level picture of a quantum system,
so it is difficult to apply the scheme to large systems.

In this short note, we propose a new approach
to obtain an analytic optimal field for complex quantum systems \cite{Haake01}.
Under a ``coarse-grained picture'' with OCT,
which is valid for such a complex system,
we derive an analytic expression for the optimal field
which steers initial states to target states in a certain limit.
By numerically solving Schr\"odinger equations,
we confirm that perfect control is actually achieved. 
This point is important because 
the zeroth-order solutions of OCT \cite{DR95,ZR99}, which look similar 
to our result, are not guaranteed to achieve perfect control. 
Another point is that our final expression does not 
require a detailed information from the energy level picture,
so that it is easy to apply to large quantum systems, in principle.

We use OCT as a theoretical vehicle.
The aim of OCT is to obtain
an optimal field $\epsilon(t)$ which guides the system
from an initial state $\left|\Phi_i\right>$ at $t=0$ to
a target state $\left|\Phi_f\right>$ at some specific time $t=T$.
According to the OCT scheme by Zhu, Botina, and Rabitz \cite{RZ00,ZBR98},
the optimal field for the Hamiltonian $H[\epsilon(t)]=H_0+\epsilon(t)V$ is
given by
\begin{equation}
\label{eqn:field}
   \epsilon(t)=\frac{1}{\alpha\hbar}
     {\rm Im}\left[
        \langle \phi(t)| \chi(t) \rangle
        \langle \chi(t)| V | \phi(t) \rangle
     \right],
\end{equation}
where the quantum state $\left|\phi(t)\right>$ and
the inversely-evolving quantum state $\left|\chi(t)\right>$ satisfy usual 
Schr\"odinger equations
with boundary conditions $|\phi(0) \rangle=|\Phi_i \rangle$ 
and $|\chi(T) \rangle=|\Phi_f \rangle$.
In order to maximize the final overlap $J_0=|\left<\phi(T)|\Phi_f\right>|^2$,
numerical iterations \cite{ZBR98} are necessary.

We {\it approximately} solve this problem 
by introducing ``coarse-grained (CG)'' Rabi states.
We extend the usual Rabi oscillation between two eigenstates
to the oscillation between two {\it time-dependent} states,
$\left|\phi_0(t)\right>$ and $\left|\chi_0(t)\right>$, defined by
\begin{equation}
\label{eqn:basis}
  \left|\phi_0(t)\right>\equiv U_0(t,0)\left|\Phi_i\right>,\quad
  \left|\chi_0(t)\right>\equiv U_0(t,T)\left|\Phi_f\right>
\end{equation}
where $U_0(t_2,t_1)$ is the propagator for $\epsilon(t)=0$ from $t_1$ to $t_2$.
We introduce the CG Rabi states as
\begin{eqnarray}
\label{eqn:forward}
  |\phi(t)\rangle
    &=& |\phi_0(t)\rangle\cos\Omega t
    -ie^{-i\theta}|\chi_0(t)\rangle\sin\Omega t,\\
\label{eqn:inverse}
  |\chi(t)\rangle
    &=&-ie^{i\theta}|\phi_0(t)\rangle\sin \Omega(t-T)\nonumber\\
    &&\qquad+|\chi_0(t)\rangle\cos\Omega(t-T).
\end{eqnarray}
The phase $\theta$ is determined from 
a normalization condition $\langle \phi(t)|\phi(t) \rangle=1$ [or
$\langle \chi(t)|\chi(t) \rangle=1$] as
\begin{equation}
e^{2 i\theta}
=\frac{\langle \phi_0(t) | \chi_0(t) \rangle}
  {\langle \chi_0(t) | \phi_0(t) \rangle}
=\frac{\langle \phi_0(T) | \Phi_f \rangle}
  {\langle \Phi_f | \phi_0(T) \rangle}
\label{eqn:phase}
\end{equation}
and $\Omega$ is a constant determined later.
It is also shown \cite{TFM04} that 
the CG Rabi states (\ref{eqn:forward}) and (\ref{eqn:inverse})
satisfy Schr\"odinger equations with the optimal field derived below
when $|\Phi_i \rangle$ and  $|\Phi_f \rangle$
contain many eigenstates without a special correlation,
i.e., they are random vectors \cite{Haake01},
and the target time $T$ is long enough.

Substituting eqs.~(\ref{eqn:forward}) and (\ref{eqn:inverse})
into eq.~(\ref{eqn:field}), and after some manipulations,
we have an expression for the optimal field,\cite{TF04,TFM04}
\begin{equation}
\label{eqn:opt-field}
  \epsilon(t)
   =\frac{\sin2 \Omega T}{2\alpha\hbar}{\rm Re}\left[
       e^{-i\theta}\left<\phi_0(t)\right|V\left|\chi_0(t)\right>
  \right].
\end{equation}
We have used $|\langle \phi_0(t)| \chi_0(t) \rangle| \ll 1$ 
which is justified when 
$|\Phi_i \rangle$ and  $|\Phi_f \rangle$
are random vectors as above.

Substituting eq.~(\ref{eqn:forward}) [or (\ref{eqn:inverse})] 
into the Schr\"odinger equation with the field (\ref{eqn:opt-field}),
and invoking the rotating-wave approximation (RWA),
i.e., omitting highly oscillating terms \cite{AE87},
we have
\begin{equation}
\label{eqn:final-omega}
  \Omega=\frac{\bar V^2\sin2 \Omega T}{4\alpha\hbar^2},
\end{equation}
where
\begin{equation}
  \bar V=\sqrt{\lim_{\tau\rightarrow\infty}\frac{1}{\tau}\int_0^\tau
     \left|\left<\phi_0 (t)\right|V\left|\chi_0(t)\right>\right|^2dt
   }
\label{eqn:v2}
\end{equation}
is an ``average transition element'' between the time-dependent
states~(\ref{eqn:basis}).
These equations constitute a set of equations
to determine the value of $\Omega$
for given $\alpha$ and $T$.
The final overlap follows as
\begin{equation}
\label{eqn:final-overlap}
  J_0=\sin^2 \Omega T,
\end{equation}
and the amplitude of the optimal field $\bar \epsilon$ is given by
\begin{equation}
\label{eqn:amplitude}
  \bar\epsilon\equiv\sqrt{\lim_{\tau\rightarrow\infty}
    \frac{1}{\tau}\int_0^\tau|\epsilon(t)|^2dt}
    =\frac{\sqrt{2}\hbar\Omega}{\bar V}.
\end{equation}

\begin{figure} 
\begin{center}
\includegraphics[scale=0.6]{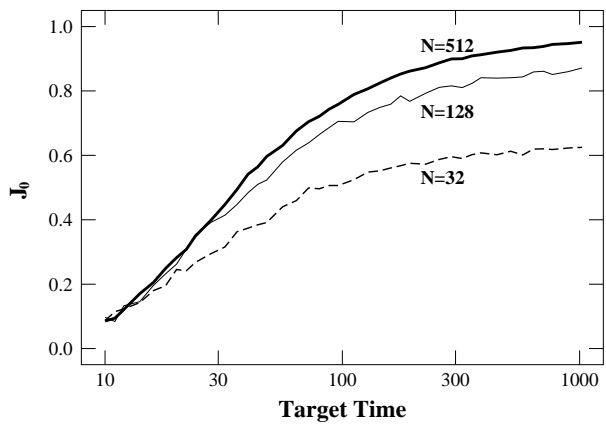}
\end{center}
\caption{
\label{fig:analytic}
The final overlap driven by the analytic optimal field,
eq.~(\ref{eqn:perfect-field}),
is shown as a function of the target time $T$.
The system is random matrices with size $N$. 
The initial and target states are random vectors 
with Gaussian random elements, and 
the average $J_0$'s for 100 ensembles are depicted.
}
\end{figure}

\begin{figure} 
\begin{center}
\includegraphics[scale=0.6]{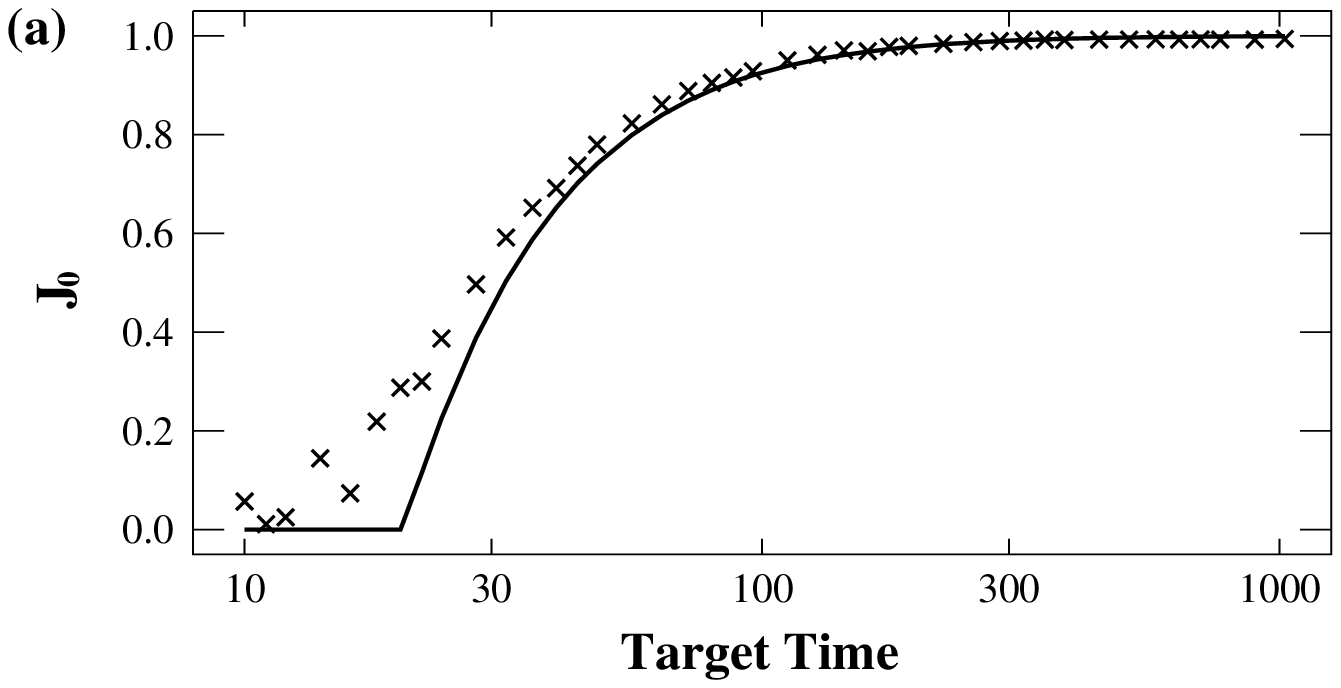}\\
\includegraphics[scale=0.6]{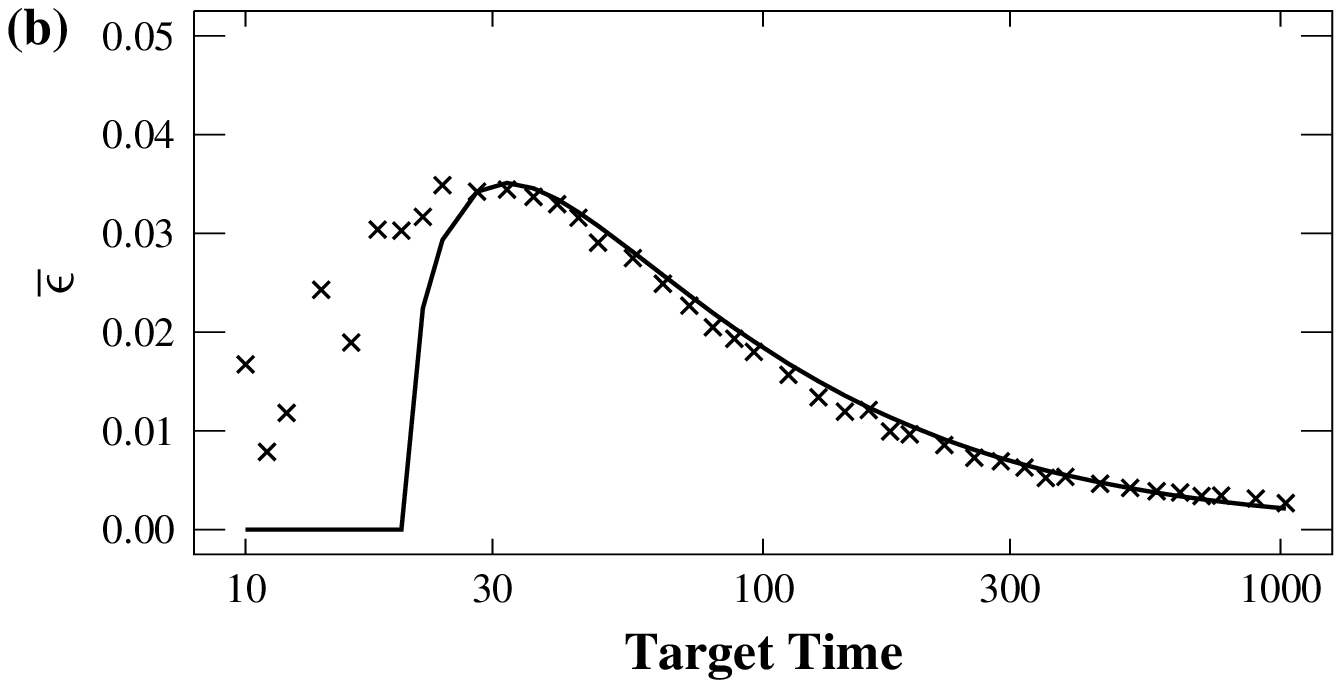}
\end{center}
\caption{
\label{fig:overlap-amplitude}
(a) The final overlap
$J_0=\left|\left.\left<\phi(T)\right|\Phi_f\right>\right|^2$ and
(b) the averaged field amplitude $\bar\epsilon$ are shown as a 
function of the target time $T$.
Marks ($\times$) represent numerical results by solving 
OCT equations using the Zhu-Botina-Rabitz scheme \cite{ZBR98}.
Solid curves represent our analytic results using the CG Rabi state. 
}
\end{figure}

Furthermore we set perfect control $J_0=1$, which is realized
when $\Omega=\pi/2T$ [See eq.~(\ref{eqn:final-overlap})].
This is equivalent to take the limit $\alpha\rightarrow0$ 
[See eq.~(\ref{eqn:final-omega})].
Substituting eq.~(\ref{eqn:final-omega}) into eq.~(\ref{eqn:opt-field}),
and using $\Omega=\pi/2T$, we obtain the final expression,
\begin{equation}
\label{eqn:perfect-field}
  \epsilon(t)=\frac{\pi\hbar}{\bar V^2T}{\rm Re}\left[
    e^{-i\theta}\left<\phi_0(t)\right|V\left|\chi_0(t)\right>
  \right].
\end{equation}
Note that this contains the same factor $\langle\phi_0(t)|V|\chi_0(t)\rangle$
as the zeroth-order solution of OCT \cite{DR95,ZR99},
but the prefactor is introduced for perfect control.
Actually the pulse area $A(t)$ can be defined by
\begin{equation}
\label{eqn:area}
  A(t)\equiv\left|\int_0^t\epsilon(t')
    \frac{2\langle\phi_0(t')|V|\chi_0(t')\rangle}{\hbar} dt'\right|,
\end{equation}
as an extension of the usual two-state case \cite{AE87}.
If we substitute eq.~(\ref{eqn:perfect-field}) into eq.~(\ref{eqn:area}),
and use eq.~(\ref{eqn:v2}), we obtain $A(T)=\pi$ under RWA,
i.e., eq.~(\ref{eqn:perfect-field}) is a generalized $\pi$ pulse.

In Fig.~\ref{fig:analytic},
we show that the analytic field (\ref{eqn:perfect-field})
actually works ($J_0 \rightarrow 1$)
for $N\times N$ random matrix systems \cite{Haake01}.
Since we have used the RWA and omitted ${\cal O}(1/\sqrt{N})$ 
terms in the derivation of eq.~(\ref{eqn:perfect-field}),
perfect control is achieved
when $T, N\rightarrow\infty$ as expected.
In Figs.~\ref{fig:overlap-amplitude}(a) and \ref{fig:overlap-amplitude}(b),
we confirm eqs.~(\ref{eqn:final-overlap}) and (\ref{eqn:amplitude})
by comparing with numerical calculations
using the Zhu-Botina-Rabitz scheme \cite{ZBR98}.
The agreement is good, especially for large $T$.

We have studied random matrix systems as ``complex'' quantum systems,
while real systems are located
between simple systems and random matrix systems.
Hence it will be interesting to apply this analytic field to more 
realistic cases of banded random matrix 
systems, i.e., a transition dipole matrix $V$ is a sparse 
matrix and there can be no direct transitions between arbitrary 
quantum states.
Our analytic field often fails to control in the situation
that single-photon processes are forbidden,
while we can show that nearly perfect control is achieved\cite{TF04}
if we restrict the initial and target state.
Even in such a case, the next order solution is easily constructed
along our picture,\cite{TF04}
and can be applicable to multi-photon processes.

The authors thank Prof.~S.A.~Rice, Prof.~H.~Rabitz, and Prof.~M.~Toda
for useful discussions.
The research of the authors has been supported in part by JSPS
Grant-in-Aid No.~14077213.

\end{document}